\documentstyle[prl,aps,epsf,twocolumn]{revtex} 
\input epsf

\def\({\left(}  
\def\){\right)}

\begin{document}
\draft
\twocolumn[\hsize\textwidth\columnwidth\hsize\csname@twocolumnfalse%
\endcsname

\preprint{}
\title{Conformal field theory and edge excitations for the principal series of 
quantum Hall fluids}
\author{V.~Pasquier and  D.~Serban}
\address{
{\it Service de Physique Th\'eorique, 
CE-Saclay, F-91191 Gif-Sur-Yvette, France.}
}

\date{\today}
\maketitle

\begin{abstract}

Motivated by recent experimental results, we reconsider  
the theory
of the edge excitations for the fractional Hall effect at filling factors
$\nu=p/(2np+1)$.
We propose to modify the standard $u(1)\otimes su(p)$ edge theory
for this series by introducing twist fields which change
the boundary conditions of the bosonic fields.
This has the effect of removing the conserved charges associated to 
the neutral modes while keeping the right statistics of the particles. 
The Green function of the electron in presence of twists
decays at long distance   
with an exponent varying continuously with $\nu$.

\end{abstract}
\vspace{0.5cm}
]

  
The low energy physics of the quantum Hall samples is governed by excitations 
located at the edge, the bulk excitations being gapped \cite{hal}. 
Recently, it became 
possible to test the nature of these excitations through experiments involving
tunneling conductance measurements
\cite{chang,gray}.
Therefore, it is important to have a reliable theoretical description of the
edge physics.
Wen \cite{wen} showed that for filling factors $\nu=1/(2n+1)$, the edge 
can be described as a chiral Luttinger liquid.
This theory was extended \cite{froh} to filling fractions of the form 
$\nu=p/(2np+1)$, where it 
involves $p$ independent chiral bosons. 
The appearance of the quantum Hall effect at these 
filling fractions is understood in terms of the hierarchical construction
\cite{hh,read}, or more intuitively using the composite 
fermions picture \cite{jain}: 
the electrons trap an even number ($2n$) of quanta of flux, see 
a reduced magnetic field and show their own  
quantum Hall effect at integer effective filling fraction $p$. 
The $p$ effective Landau levels are thought to be at the origin 
of the $p$ components of the corresponding field theory \cite{wenzee}. 
It was shown \cite{froh} that the edge theory is a conformal field theory
with affine symmetry
$u(1)\otimes su(p)$. 
A different version of the conformal field theory
on the edge, based on the $W_{1+\infty}$, was proposed 
\cite{cappelli}.
     
The experiments \cite{chang,gray} have tested the Luttinger liquid nature 
of the edge by measuring the tunneling conductance between 
an electron gas and a fractional quantum Hall fluid, and found a power  
law dependence on the voltage, $dI/dV \sim V^ {\alpha-1}$.  
Surprisingly, the exponent $\alpha$ varies linearly with $1/\nu$,  
instead of being   
discontinuous, as predicted by the theory \cite{kf} .

Several attempts were made to conciliate the 
predictions of the edge theory with the experiments  
\cite{halperin,zuma,lw,flop}.  
The basic problem to be solved concerns the electron's Green function.  
The electron is a fermion, so the Green function has  
to be odd under the exchange of the fermion coordinates. On the other  
hand, to explain the observed dependence of the conductance, the  
Green function should behave as $t^{-1/\nu}$ at large time separations.
The two requirements seem to be contradictory, at least in the 
context of conformal field theory for the edge.   
The second requirement is fulfilled if only the mode carrying 
the electric charge  
propagates. Different mechanisms were proposed to suppress the contribution
of the neutral modes,  
otherwise necessary to reestablish the fermionic electron statistics.    
 
In this paper, we show that a consistent mechanism for the suppression of 
the neutral modes exists in the framework of conformal field
theory. We consider a modification  
the standard $p$-boson conformal field theory \cite{froh}  
by introducing twist operators. They create branch cuts
for the chiral boson fields, change their boundary conditions,
and therefore interpolate between sectors of the Hilbert space
with different quasi-periodicity properties. 
Twist operators 
were considered in the context of conformal field  
theory on Riemann surfaces \cite{kn,ising}  and string theory on orbifolds 
\cite{br}.
The first effect of twists 
is to break the global $su(p)$ symmetry,  
therefore removing the charges associated to the neutral sector, 
the $su(p)$ spin index being replaced by a branch index.
Second, the presence 
of twist insertions in the electron Green function modify their behavior,
depending on the twist coordinates.
If a twist is introduced (removed) at $t=-\infty$ $(t=\infty)$, then, 
in the thermodynamic limit we retrieve the same behavior of the 
Green function as in ref. \cite{kf}. 
In the limiting case where twists
are glued to the  electrons, the neutral modes are completely suppressed
and the Green function behaves as  
$t^{-1/\nu}$ at time separation $t$.
  
To trace back the significance of the twists,
in the last section we investigate the properties of the bulk wave   
functions of free electrons  
in magnetic field at filling factor $\nu=p/(2np+1)$.   
Since at this filling factor there are $2n+1/p$ flux quanta per  
particle, the quantization of the system in a finite geometry
(on the torus, for example)   
can be done consistently only by considering  
$p$-valued wave functions. 
The Hilbert space splits into $p$ quasi-periodicity sectors,
which will survive even after introducing an  
interaction potential.
Reducing the electron density
to a one-dimensional density on the edge, we retrieve 
a bosonic fields with twisted boundary conditions.
We are lead to conclude that the presence of the twists in the edge field theory
reflect the particular structure of the  Hilbert space at 
the filling factors $\nu=p/(2np+1)$.

{\it Conformal Field Theory for twisted bosons.}  
Let us first briefly present the basic ingredients of the 
standard theory of the edge excitations, from the point of view 
of a conformal field theory \cite{dif}. 
We use the complex coordinate $z=e^{t-ix}$, in which the surfaces of  
equal times are circles (radial quantization). 
Since the excitations at one edge are chiral, we 
consider only the holomorphic sector of the theory. 
 
Define $p$ bosonic fields by : 
	\begin{eqnarray} 
	\label{notwist} 
	\varphi_k(z)=\varphi_0^{(k)}&-&ia_0^{(k)} \ln{z}+ 
	i\sum_{m\neq 0, m \in {\bf Z}}   
	\frac{a_m^{(k)} z^{-m}}{m} \;,\ \ {\rm{with}}\cr  
	\noalign{\vskip3pt}
	[a_m^{(k)},a_{m'}^{(l)}]&=&
	m \delta_{m+m',0}\;
	\delta_{k,l} \;,    
	\quad [\varphi_0^{(k)}, a_0^{(l)}]=i \delta_{k,l}\;. 
	\end{eqnarray} 
The bosonic modes 
$a_m^{(k)}$ with $m<0$ play the role of creation operators. 
The fundamental object insuring the conformal invariance is the  
stress-energy tensor: 
	$$ T(z)=-\frac{1}{2}\sum_{l=0}^{p-1}:(\partial   
	\varphi_l(z))^2:\;  
	$$ 
with conformal anomaly, or central charge $c=p$. The double dots 
represent normal ordering with respect to the bosonic operators.   
There are $p$ conserved 
charges $a_0^{(k)}$, corresponding to the conserved currents  
$\partial \varphi_k(z)$.

The charged excitations in the spectrum, electrons and quasi-particles,  
are created by exponentials of bosonic fields (vertex operators). 
Electrons and quasi-particles are characterized by the dual parameters
$\beta$ and $\beta'$, defined by:
	$$ \beta=2n\;,\qquad p\beta'+1={1}/({p\beta+1})\;.
	$$  
For the standard theory of edge excitations \cite{wenzee}, 
the electron and quasi-particle operators are:  
	$$
	\Psi^{(k)}_e(z)=
	:e^{i \sum_j v^k_j \varphi_j(z)}:\;, \ \   
	\Psi^{(k)}_q(z)=:e^{i \sum_j s^k_j \varphi_j(z)}:\;
	$$  
with $\sum_{j=0}^{p-1} v^k_j  v^l_j= \beta +\delta_{kl}\equiv K_{kl}$,  
$\sum_{j=0}^{p-1} s^k_j  v^l_j= \delta_{kl}$ 
and $\sum_{j=0}^{p-1} s^k_j  s^l_j= \beta'+\delta_{kl}= (K^{-1})_{kl}$. The vectors   
$v^k$ generate the $p$ dimensional charge lattice of a $u(1) \otimes su(p)$ 
algebra, characterized by the $p \times  p$ matrix $K$ \cite{read,wenzee} 
and the vectors $s^k$  generate the dual lattice. 
The index $(k)$ of the vertex operators is a  
$su(p)$ spin index. 
Properties as the statistics or the dimension of the operators  
can be deduced from the product of two operators : 
	$$  
	\Psi^{(k)}_e(z) \Psi^{(l)}_e(w)=  
	(z-w)^{\beta+\delta_{kl}}:\Psi^{(k)}_e(z) 
	\Psi^{(l)}_e(w):\;. 
	$$
The electron Green function can also be determined :
	$$ \langle \Psi^{(k)\dagger}_e(z) \Psi^{(k)}_e(w)
	\rangle=(z-w)^{-\beta-1}\;,
	$$
leading to the exponent for the conductance $\alpha=\beta+1$.
Quasi-particle have similar properties, with $\beta$ replaced by $\beta'$.

Let us now allow for the presence of twists, changing the boundary
conditions of the bosonic fields. Twists appeared in the conformal field
theory in connection with the Ising model \cite{ising}, where they 
represent the 
spin operators, and in string theory \cite{kn,br}, where they were used to 
introduce branch point singularities in the background.
Here, we use $p-1$ twist operators $\sigma_1,\ldots \sigma_{p-1}$ which 
introduce branch points of order $p$
for the bosonic fields \cite{br}. We make the change of variable:  
 	$$ \tilde \varphi_{l}(z)=\frac{1}{\sqrt{p}}
	\sum_k \omega_p^{lk} \varphi_{k}(z)\;,$$
with $\omega_p=e^{2\pi i /p}$.
In terms of these fields, the monodromy conditions around the 
singularities introduced by the twists 
are given by :
	$$\partial \tilde \varphi_{l}(z) \sigma_k(0) = z^{-\alpha/p} 
	\tau_{l,k}(0)+\ldots\;,$$
where $\alpha=lk\ {\rm mod}\ p$, the dots indicate less singular terms
and the fields $\tau_{l,k}(z)$ are called excited twist fields.
The twists themselves obey the following algebra:
	$$\sigma_k(z) \sigma_l(0)=(z-w)^\delta \sigma_r(0)+\ldots\;,$$
with $r=k+l\ ({\rm mod}\ p)$, $\sigma_0\equiv 1$ and $\delta$ is a number which can be 
determined 
by dimensional analysis. A twist operator interpolates between
two different sectors of the Hilbert space.
 
First, we will consider the simpler case when there is one twist $\sigma_1$
located at $z=0$ and one $\sigma_{p-1}$ located at $z=\infty$
(the global twist should be equal to zero).
The vacuum in this twisted sector is
$|\sigma_1\rangle=\sigma_1(0)|0\rangle$, on which  
the $p$ bosonic fields have the following mode decomposition :  
 	\begin{eqnarray}  
 	\label{bosons}  
	&\partial \tilde \varphi_l(z)&=-i\sum_{m \in {\bf Z}+l/p}   
	{a_m z^{-m-1}} \;, \cr \noalign{\vskip5pt}  
	&{\rm{with}}& \qquad [a_m,a_{m'}]=m\delta_{m+m',0}\;,   	   
	\end{eqnarray}  
Here, the bosonic modes $a_m$ are indexed by a fractional  
number $m\in {\bf Z}/p$, the ones with $m<0$ 
representing the creation operators. 
Note that the commutation relations 
naturally couple the field $\tilde\varphi_l(z)$ to $\tilde \varphi_{p-l}(z)$. 

The holomorphic component of the stress energy tensor  
in the sector twisted by $\sigma_1$ is:  
	\begin{eqnarray} 	
	\label{stress} 
	\tilde T(z)=-\frac{1}{2}
	&:&\big((\partial \tilde \varphi_0(z))^2+  
	\partial\tilde \varphi_1(z)\partial \tilde \varphi_{p-1}(z)+\ldots +\cr  
	&+&  
	\partial \tilde \varphi_{p-1}(z)\partial \tilde \varphi_1(z)\big):+  
	\;\frac{p^2-1}{24p z^2}\;.  
	\end{eqnarray}  
Note that the stress energy tensor is single valued around $z=0$, although  
the constituting bosons are not.  
The last term in $T(z)$ gives  
the dimension of the  twist operator $\sigma_1$, $\Delta_1= ({p^2-1})/{24p}$.
Similarly, one obtains for the twisted sectors
generated by $\sigma_k$,  $\Delta_k= ({p^2-q^2})/{24p}$,
where $q$ is the greatest common divisor of $p$ and $k$ \cite{br}.
 
From the mode decomposition (\ref{bosons}), we deduce that 
there is no conserved charge associated to the twisted bosons
$\tilde\varphi_1,\ldots, \tilde\varphi_{p-1}$. 
Even if there are conserved currents associated to the twisted sector, 
they {\it do not} manifest at the level of  global symmetries. 
The only conserved charge is the electric charge 
$J_0=\sqrt{\nu} a_0$, normalized  
such that the electron charge is $1$.

{\it Electron and quasi-particle operators in presence of twist.}  
In the
presence of the twist $\sigma_1(0)$
the bosonized form of the electron and the quasi-particle operators  
is the following:
	\begin{equation}
	\Psi_{e,q}^{(k)}(z)\; \sigma_1(0)=z^{-\gamma}\;
	\tilde \Psi_{e,q}^{(k)}(z)\; \sigma_1(0)\;, 
	\end{equation}
with:   
	\begin{eqnarray}   
	\label{eqp}  
	\tilde\Psi_e^{(k)}(z)&=& :e^{\frac{i}{\sqrt{p}} 
	\left(\sqrt{p\beta+1}\tilde\varphi_0(z)   
	+\sum_{j=1}^{p-1}\omega_p^{kj}\;  
	\tilde\varphi_j(z)\right)} :\;, \cr   
	\tilde\Psi_q^{(k)} (z)&=& :e^{\frac{i}{\sqrt{p}}  
	\left(\frac{1}{\sqrt{p\beta+1}}\tilde\varphi_0(z) +
	\sum_{j=1}^{p-1}\omega_p^{kj}\;  
	\tilde\varphi_j(z)\right)} :\;.    
	\end{eqnarray}
The power $\gamma$ can be computed by dimensional analysis, 
in the following way. Let $\Delta_e=(\beta+1)/2$ and
$\Delta_q=(1-\beta/(p\beta+1))/2$ be the dimensions 
of the operators $ \Psi_{e,q}^{(k)}(z)$
respectively. In the presence of the twist $\sigma_1$, the vertex operators
$\tilde \Psi_{e,q}^{(k)}(z)$ have the dimensions 
$\tilde \Delta_{e}=1/2\nu$ and $\tilde \Delta_{q}=\nu/2p^2$.
The reason for this change in the dimensions of the vertex operators
is the absence of the zero modes of the fields $\tilde\varphi_1,
\ldots,\tilde\varphi_{p-1}$.
We obtain
$\gamma=\Delta_e-\tilde \Delta_e=\Delta_q-\tilde \Delta_q=\(1-
{1}/{p}\)/2$.

As the bosons (\ref{bosons}),  
the vertex operators have a non-trivial monodromy around the point 
$z=0$:    
	$$ z \to z e^{2i\pi } \quad \Rightarrow \quad
	\Psi_e^{(k)}(z )\to c_k \Psi_e^{(k-1)}(z)\;.  
	$$ 
where $c_k$ is a phase factor, and similarly for $ \Psi_q^{(k)} (z)$. 
The index 
$(k)$ has to be understood modulo $p$, and it represents 
now a branch index rather than a spin index.  
Since the $p$ operators $\Psi_e^{(k)} (z)$ represent different determinations  
of the same operator, we will drop the branch index 
in the following. 

Next, we are going to compute the Green function of the electron operator,
in the presence of two twists at arbitrary positions $z_1$, $z_2$:
	$$
	G(z,w;z_1,z_2)=\frac{\langle \sigma_{p-1}(z_1)\; \Psi_e^\dagger(z)
	\; \Psi_e(w)\; \sigma_1(z_2)\rangle}
	{\langle \sigma_{p-1}(z_1)\; \sigma_1(z_2)\rangle}
	$$
The short distance $(z\to w)$ singularity of $G(z,w;z_1,z_2)$
is given by $(z-w)^{-\beta-1}$. From the behavior of the
correlation function under conformal transformations, we have:
	$$
	G(z,w;z_1,z_2)=(z-w)^{-\beta-1} F(x)\;, \  
	x=\frac{(z-z_1)(w-z_2)}{(z-z_2)(w-z_1)}
	$$ 
with $F(x)$ some function of the cross ratio $x$, which can be determined
in the limit $z_1 \to \infty$, $z_2 \to 0$ by using the bosonization
(\ref{eqp}):
	$$F(x)=x^{-\frac{1}{2}\(1-\frac{1}{p}\)}\frac{x-1}{x^{1/p}-1}\;.
	$$
Since $F(x)=F(1/x)$, the Green function $G(z,w;z_1,z_2)$ is odd under
exchange of $z$ and $w$. 
It is possible to remove the short distance singularity
in the product of electron and twist operator, allowing to bring them
eventually to the same point:
	$$:\Psi_e(z)\; \sigma_1(w):\; \equiv
	(z-w)^\gamma \Psi_e(z)\; \sigma_1(w)\;.
	$$
Bringing the electron and the twist together cancels the neutral modes, as can be seen 
from the corresponding Green function:
	\begin{eqnarray*}
	G'(z,w;z,w)&=&\frac{\langle :\sigma_{p-1}(z)\; \Psi_e^\dagger(z):
	\;: \Psi_e(w)\; \sigma_1(w):\rangle}
	{\langle \sigma_{p-1}(z)\; \sigma_1(w)\rangle} \cr
	\noalign{\vskip5pt}
	&=&(z-w)^{-\beta-1/p}=(z-w)^{-1/\nu}\;.
	\end{eqnarray*}
We infer that this Green function correspond to the insertion of charge only
into the system.
If the same mechanism is applied to the 
quasi-particle excitations, the exponent for the quasi-particle 
Green function will be given by $Q^2/\nu=\nu/p^2$.

For comparison, if the twists are located at $z_1 = \infty$, $z_2 = 0$
(corresponding to $t_1=\infty$, $t_2=-\infty$), the Green function is:
	\begin{eqnarray*}
	G'(z,w;\infty,0)&=&\frac{\langle :\sigma_{p-1}(\infty)\; 
	\Psi_e^\dagger(z):
	\;: \Psi_e(w)\; \sigma_1(0):\rangle}
	{\langle \sigma_{p-1}(\infty)\; \sigma_1(0)\rangle} \cr
	\noalign{\vskip5pt}
	&=&(z-w)^{-\beta}(z^{1/p}-w^{1/p})^{-1}\;.
	\end{eqnarray*}
In the thermodynamic limit $t\ll L/v\equiv 2\pi$, where $t$ is the 
time separation between the points
$z$ and $w$,  this  last Green function behaves like
$t^{-\beta-1}$ and the effect of the twist cannot be detected.

{\it Electrons on the torus at $\nu=p/(p\beta+1)$.}  
In order to understand better the origin of the twists  
in the edge theory we go  
back and study the problem of $N$ electrons at filling   
factor $\nu=p/(p\beta+1)$. 
For the sake of simplicity, we use a torus 
of periods $L_x$ and $L_y$ and area $S=L_xL_y$,

In the absence of Coulomb interactions, the Hamiltonian is a sum of 
one-particle terms of the type:  
	$$H_1=\frac{1}{2m_e}\(\vec p-e\vec A\)^2\; $$  
The magnetic translations $  
T(\vec R)=e^{\frac{i}{\hbar}\vec R (\vec p +e\vec A)}$  
commute with the Hamiltonian, but do not mutually commute.    
The magnetic translations along the two  
periods of the torus, $T_x$ and $T_y$ obey the commutation  
relation:  
	$$T_x T_y =e^{2\pi i \frac{BS}{\phi_0} }T_y T_x\;,$$  
$\phi_0=h/e$ being the flux  
quantum. The number of flux quanta through the system
$BS/\phi_0=N/\nu=N(\beta+1/p)$ is in general fractional, so that
the two translations cannot be diagonalized simultaneously
and one cannot find a basis of periodic one-particle wave functions.
Put differently, the number of  states per Landau level, 
$N/\nu=N(\beta+1/p)$,  should be fractional.
Since $T_x^pT_y=T_yT_x^p$, the wave functions could be rendered periodic on  
a $p$-cover of the torus, with periods $pL_x$ and $L_y$ for example.
To preserve the filling factor, we have to consider all the wave functions 
periodic on the $p$-cover of the initial torus and fill them with 
$pN$ electrons.


Let us  
construct explicitly the one-particle wave functions on the torus  
from the wave functions on the plane, in the gauge $A_x=-By$,  
$A_y=0$. For $b=eB/\hbar>0$, the wave functions  
in the lowest Landau level are:  
	$$\phi_m(x,y) = C\; e^{-b\left(y-{2\pi m}/{bL_x}  
	\right)^2/2}  
	e^{-{2\pi i mx}/{L_x}}\;,$$  
with $C$ a normalization constant.  
To render these functions periodic over $L_y$, one takes:  
	$$\Phi_m(x,y)=\sum_{M\in {\bf Z}}\phi_{m+MN(\beta+1/p)}(x,y)\;.$$   	  
If $m \in {\bf Z}/p$, the resulting function  
will be periodic over $pL_x$. 
Their behavior under translations by $L_x$ depend on the value of $N$.
We identify $p$ sectors in the Hilbert space, indexed by $k=0,\ldots,p-1$, 
where the number of electrons is $N=k\ {\rm mod}\ p$.
For simplicity, we consider in the following the sector $k=1$. 
Then, the wave functions $\Phi_m(x,y)$ decompose into $p$ sectors: 
	\begin{eqnarray*}  
	\Phi_{m}(x,y)&=&\sum_{j=0}^{p-1} \Phi_{m}^{(j)}(x,y)\;, \cr   
	\Phi_{m}^{(j)}(x+L_x,y)&=&\omega_p^j \; \Phi_{m}^{(j)}(x,y)\;.  
	\end{eqnarray*}  
With $m=1/p,2/p,...,N/\nu$, they  constitute a  
$N(p\beta+1)$ dimensional  basis of the lowest Landau level
on the torus of area $pL_xL_y$.
A small perturbation potential, periodic on the initial
torus, will preserve the quasi-periodicity sectors, since the
matrix elements of the potential between two different 
periodicity sectors
vanish.

To deduce the implications of the quasi-periodicity
for the edge excitations, we consider the situation   
where two (sharp) edges are created at $y=0$ and $y=L$,  
by introducing some confining potential. For simplicity, we take $L,L_y\to \infty$   
so that the wave functions are equal to $\phi_m(x,y)$, normalized by   
$C=(pL_x \sqrt{\pi/b})^{-1/2}$.   
The electron operator reads:  
	$$\Psi(x,y)=\sum_{m\in {\bf Z}/p}\phi_m(x,y) \; c_m\;.$$  
Following the procedure used in ref. \cite{zuma}, we define a one-dimensional   
density operator, by integrating the two dimensional density operator  
from $y=-\infty$ to some cutoff $y=Y$ in the bulk:  
	$$\rho(x)=\int_{-\infty}^{Y} \Psi^\dagger(x,y)\; \Psi(x,y)\; dy\;.$$   
We obtain a one-dimensional  
density operator which is not periodic in $x$:  
	\begin{eqnarray*}  
	\rho (x)=\sum_{l=0}^{p-1} \rho^{(l)} (x)\;&,& \quad   
	\rho^{(l)} (x+L_x)=\omega_p^l \; \rho^{(l)} (x)\;, \cr  
	\rho^{(l)} (x)&=&\sum_{m\in {\bf Z}+l/p} e^{2\pi imx/L_x} \rho_m\;.  
	\end{eqnarray*}  
Replacing the fermion number with its   
average in the ground state,  
$\langle 0 | c_m^\dagger c_{m'}|0\rangle =\nu \; \delta_{m,m'}$,  
one can show that  
the low modes of the one-dimensional density operator, $|m| \ll L_x\sqrt b/2\pi$,   
obey bosonic commutation relations:  
	$$[\rho_m,\rho_{m'}]=\nu\; m \;\delta_{m+m',0}\;, \quad  
	m,m'\in {\bf Z}/p\;.$$  
With $\rho_m$ replaced by $\sqrt{\nu}a_m$,   
these are the commutation relations of the bosonic fields in the sector twisted 
by $\sigma_1$   
(\ref{bosons}). By this identification,   
$\rho_0=\sqrt{\nu} a_0$  
represents the charge operator. Compared to Z\"ulicke and MacDonald  
\cite{zuma}, we obtain non-integer modes for the density operator.  
They correspond to the neutral modes which 
restore the fermionic statistics   
of the electron operator.

In conclusion, we have shown that the requirement that only the charge
mode contributes to the tunneling conductance can be formulated in
in a consistent way in the language of conformal field theory. 
Our conclusions essentially agree
with that of references \cite{zuma,lw,flop} , the difference coming
from the treatment of the neutral modes. 
We argue that
at filling fraction $\nu=p/(2np+1)$, there are
$2n+1/p$ flux quanta per electron, and
the structure of the Hilbert space for the bulk electrons 
induced by the nonintegrality of the flux
is translated on the edge by the help of the twist operators.
The twists are able to account for the absence of the
neutral modes contribution from the correlation functions, 
being an alternative to setting the neutral mode velocity to zero
from the beginning \cite{lw} .
More insight about the nature of the quasi-particle excitations 
come from an analysis of the bulk. It is possible to obtain, in a 
Landau-Ginsburg-Chern-Simons approach for the bulk, vortex-like solutions 
which carry
only one charge and show a multivalued structure 
reminiscent from twists \cite{vortex}. 

As a technical remark, we point out that the relation between the conformal field 
theory with $u(1)\otimes su(p)$ symmetry and its twisted version
borrows a lot of similarity to the relation between two integrable models,
namely the Calogero-Sutherland model with $su(p)$ symmetry
and a certain limit of the Ruijenaars-Schneider model, relation
uncovered by Uglov \cite{uglov} . The bosonization of the charged
excitations in the twisted sectors is suggested by this correspondence
(see also \cite{mb} ).
 
We would like to thank M. Bershadski, A. Cappelli, E. Fradkin, 
C. Glattli, K. Imura and N. Sandler for useful discussions.  
After completing this work, we became aware of ref. \cite{maiella}
where twist operators were also used to determine the ground state 
wave function.

\end{document}